\documentclass[12pt,preprint]{aastex}

\catcode`\@=11
\newcommand{\gapprox}{\mathrel{\mathpalette\@versim>}}
\newcommand{\lapprox}{\mathrel{\mathpalette\@versim<}}
\newcommand{\propapprox}{\mathrel{\mathpalette\@versim\propto}}
\newcommand{\@versim}[2]
  {\lower3.1truept\vbox{\baselineskip0pt\lineskip0.5truept
\ialign{$\m@th#1\hfil##\hfil$\crcr#2\crcr\sim\crcr}}}
\catcode`\@=12

\shorttitle{X-RAYS FROM KEPLER CSM}
\begin{document}

\title{X-ray Emission from Strongly Asymmetric Circumstellar Material
in the Remnant of Kepler's Supernova}

\author{Mary T. Burkey,\altaffilmark{1}
Stephen P. Reynolds,\altaffilmark{1}
Kazimierz J. Borkowski,\altaffilmark{1}
and John M.~Blondin\altaffilmark{1}}

\altaffiltext{1}{Department of Physics, North Carolina State University,
  Raleigh NC 27695-8202; reynolds@ncsu.edu}

\begin{abstract}

Kepler's supernova remnant resulted from a thermonuclear explosion,
but is interacting with circumstellar material (CSM) lost from the
progenitor system.  We describe a statistical technique for isolating
X-ray emission due to CSM from that due to shocked ejecta.  Shocked CSM
coincides well in position with 24 $\mu$m emission seen by {\sl
Spitzer}.  We find most CSM to be distributed along the bright north
rim, but substantial concentrations are also found projected against
the center of the remnant, roughly along a diameter with position
angle $\sim 100^\circ$.  We interpret this as evidence for a disk
distribution of CSM before the SN, with the line of sight to the
observer roughly in the disk plane.  We present 2-D hydrodynamic
simulations of this scenario, in qualitative agreement with the
observed CSM morphology.  Our observations require Kepler to have
originated in a close binary system with an AGB star companion.

\end{abstract}

\keywords{ISM: individual objects (\objectname{G4.5+6.8}) ---
ISM: supernova remnants --- X-rays: ISM --- supernovae: general}

\section{Introduction}

The remnant of Kepler's supernova of 1604 (``Kepler'' henceforth) has
defied classification since its optical recovery in 1943 (Baade 1943),
because of its awkward combination of clear evidence for dense
circumstellar material (originally identified through an optical
spectrum indicating enhanced N; Minkowski 1943), suggesting a
core-collapse supernova, and light curve and location (470 kpc above
the Galactic plane, for a distance of 4 kpc; Sankrit et al.~2005)
indicating a Type Ia origin.  X-ray studies with ASCA indicated a
large mass of Fe (Kinugasa \& Tsunemi 1999), and finally detailed
X-ray studies with {\sl Chandra} (Reynolds et al.~2007; Patnaude et
al.~2012) showed conclusively that the event must have been a
thermonuclear explosion (though its spectrum at maximum light may or
may not have resembled a traditional Ia event).

But the evidence for circumstellar material (CSM) has not gone away.
Since the question of the progenitor systems of SNe Ia is unsettled at
this time, characterization of the CSM is of great importance.
Observations of Type Ia supernovae consistently fail to show evidence
for CSM (except in rare cases such as 2002ic, Hamuy et al.~2003, and
2005gj, Aldering et al.~2006), a finding used as evidence in favor of
binary white-dwarf progenitor systems (double-degenerate, DD).
Single-degenerate (SD) scenarios may involve either a main-sequence or
an evolved companion (see, for example, Branch 1995; Hillebrandt \&
Niemeyer 2000), but searches for a surviving companion have for the
most part been negative (e.g., Schaefer \& Pagnotta 2012).  Just as
finding a companion would demand a SD progenitor, identifying CSM in a
Type Ia system requires an SD scenario.  However, various
possibilities for the companion star are still possible.  The most
likely suggestion is an evolved star with a slow and massive
wind (Vel\'azquez et al.~2006; Blair et al.~2007; Chiotellis, Schure,
\& Vink 2012; Williams et al.~2012), i.e., an Asymptotic Giant Branch
(AGB) star.  In a close binary, such a star would be likely to produce
an asymmetric wind, primarily in the orbital plane.  Thus
characterizing the amount and spatial distribution of CSM in the
remnant of a Type Ia supernova has important implications for the
nature of Type Ia progenitors.  Here we focus on the spatial
distribution of CSM, which we identify using a powerful statistical
technique applied to the long {\sl Chandra} observation of Kepler.

The excess of nitrogen in optical observations suggests mass loss from
an evolved star.  Various authors have performed hydrodynamic modeling
of a system with substantial mass loss moving through the ISM with
speeds of hundreds of km s$^{-1}$ (Borkowski et al.~1992; Vel\'azquez
et al.~2006; Chiotellis et al.~2012).  This modeling is able to
describe the observed dense shell to the north, but in these cases the
wind was assumed to be isotropic.  We shall show that our spatial
localization of CSM requires an asymmetric wind.

\section{Observations} 

Kepler was observed for 741 ks with the {\sl Chandra} X-ray
Observatory ACIS-S CCD camera (S3 chip) between April and July 2006.
Data were processed using CIAO v3.4 and calibrated using CALDB v3.1.0.
A large background region to the north of the remnant (covering most of
the remaining area on the S3 chip) was used for all spectra.  Spectral
analysis was performed with XSPEC v.12 \citep{arnaud96}. We used the
nonequilibrium-ionization (NEI) v2.0 thermal models, based on the
APEC/APED spectral codes \citep{smith01} and augmented by the addition
of inner-shell processes \citep{badenes06}.  There are a total of
about $3 \times 10^7$ source counts, with fewer than 3\% background
(though some of those may be dust-scattered source photons).

\begin{figure}
\centering
%\epsscale{1.15}
\epsscale{1.1}
%\plotone{f1p.png}
\includegraphics[width=3truein]{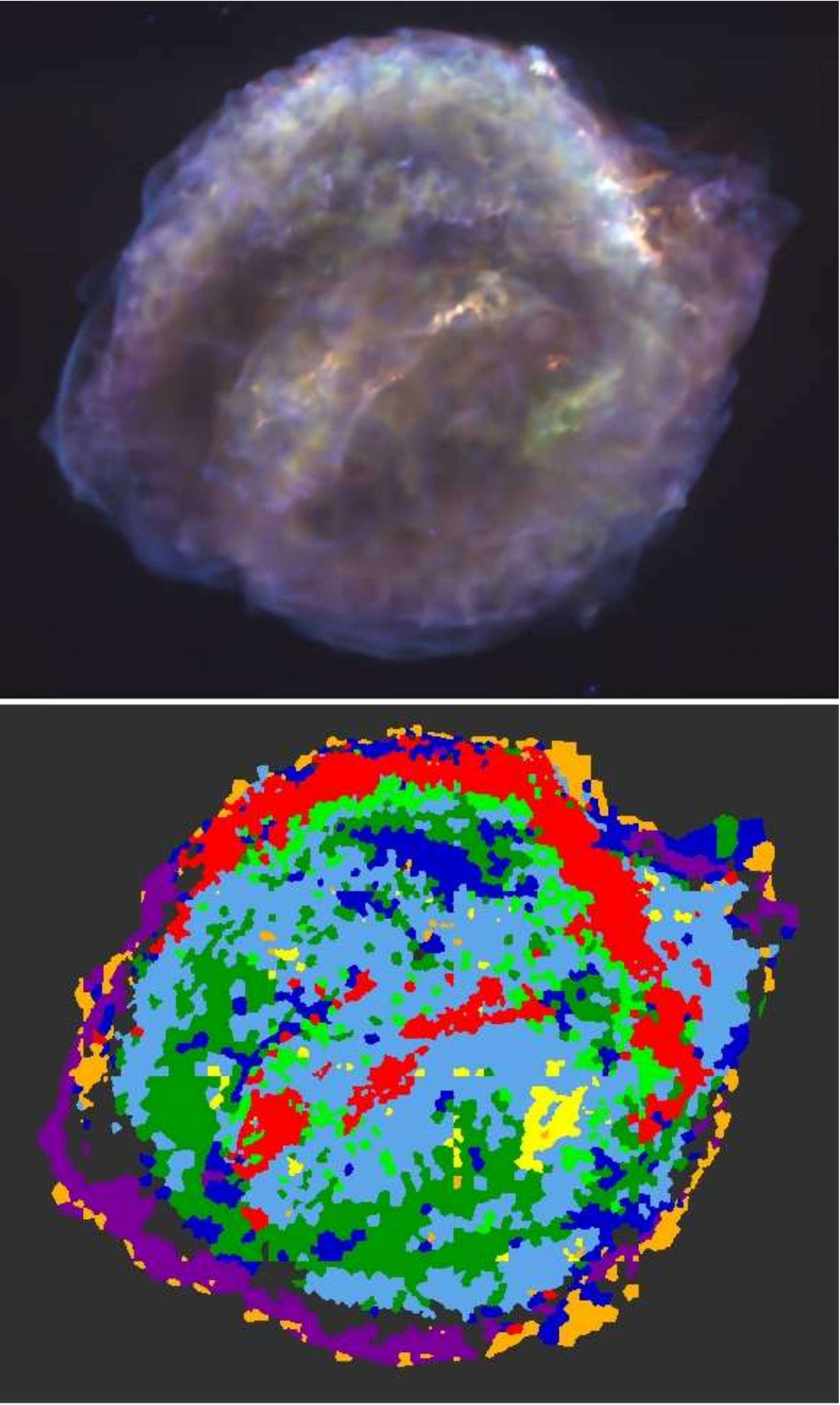}
\caption{Top: Merged image between 0.3 and 8 keV (Reynolds et
al.~2007).  Red: 0.3 -- 0.72 keV; green, 0.72 -- 1.7 keV; blue, 1.7 --
8 keV.  All three images were smoothed using platelet smoothing
\citep{willett07}.  Image size is $4\farcm7 \times 3\farcm9$.  Bottom:
Division of Kepler into clusters (regions of similar spectral character). 
Colors are arbitrary.  Region R (red) contains the bulk of
the CSM, while purple (P) is dominated by synchrotron emission.  Most of
the rest is ejecta.  See text for further details.}
%Regions from which spectra of Figs.~\ref{spectra1} and \ref{spectra2}
%\vspace{-0.1truein}
\label{ims1}
\end{figure}

\section{Gaussian Mixture decomposition}

Our earlier study (Reynolds et al.~2007) showed that spectral
variations occur over arcsecond scales, too small for full spectra of
individual regions to have adequate statistics.  This also means that
simple imaging in different spectral bands is incapable of producing
quantitative information characterizing regions of different spectral
character.  In order to concentrate regions with similar spectra for
detailed spectral analysis, we instead collect regions of similar
spectral character by describing each with four broad-band colors, and
clustering them in a four-dimensional color space using a collection
of Gaussian probability distributions (clusters). (Such a
probabilistic description of multi-dimensional data is known in the
literature as a Gaussian mixture model.)  The clusters are now of much
higher signal-to-noise than individual small regions (selected by eye,
for instance, as in Reynolds et al.~2007), and were selected with
objective criteria.  Gaussian mixture models are well-known in many
areas of science, and are becoming more widely used for a variety of
applications in astrophysics (e.g., Shang \& Oh 2012; Lee et al.~2012;
Hurley et al.~2012).

The integrated spectrum of Kepler is dominated by ejecta emission from
Fe, Si, and S, while O and Mg are expected to be primarily found in
CSM.  We therefore selected spectral bands to feature the oxygen
Ly$\alpha$ line (0.3 to 0.72 keV), the iron L shell region (0.72 to
1.3 keV), the magnesium K$\alpha$ line (1.3 keV to 1.5 keV), and the
silicon and sulfur K$\alpha$ lines (combined because of similar
characteristics and behavior in the spectrum: 1.7 to 2.1 keV and 2.3
to 2.7 keV respectively). Other elements can contribute in these
bands, for instance Ne K$\alpha$ and Ly$\alpha$ in the Fe L region,
but the integrated spectra show that these contributions are less
significant.  Each of the previous energy bands was divided by the
flux in unused energy bands (1.5 to 1.7, 2.1 to 2.3, and 2.7 to 7.0
keV), so that the entire spectral range was used.  This gave the best
signal-to-noise ratio for those fluxes, but at the cost of including
some line emission in the quasi-continuum bands.  However, the
quasi-continuum is broad enough to allow maximum contrast with the
line features we chose to highlight.  (Insisting on using a more
line-free continuum, such as 4--6 keV, has two drawbacks: first, since
far fewer counts are present there, the ratios we seek to classify
would have much higher statistical noise; and second, one cluster in
particular that we identify, that of synchrotron emission, is actually
less evident in 4--6 keV emission than in broader bands.)  The strong
contrast in the clusters we identify, as seen in Figure 2, shows that
we are not unduly hampered by the presence of some line emission in
our quasi-continuum band.

We divided the observation into about 5000 segments with sizes
adjusted to produce comparable numbers of counts (about 6000 each),
and extracted the counts in the four line bands plus continuum for
each.  We then produced a four-dimensional scatterplot of all 5000
segments using the logs of normalized band counts.  Using the publicly
available software package {\sl mclust} written in R (Fraley et
al.~2012), we decomposed the four-dimensional distribution into
several four-dimensional Gaussian clusters, each characterized by at
most 15 parameters: its weight relative to other clusters, a center in
each of four coordinates, and a (symmetric) $4 \times 4$ covariance
matrix.  The number of Gaussians required was found by the algorithm
using the Bayesian Information Criterion (BIC), which was optimized
for nine clusters.  However, the clusters corresponding to CSM emerged
clearly for any total number of clusters between five and ten.

In the Gaussian mixture model, each segment is associated with a set
of probabilities of belonging to each of the clusters.  We will from
now on refer to each cluster more narrowly as a set of segments that
have the highest probability of belonging to this cluster.
Figure~\ref{ims1} shows the spatial distribution of the segments
making up each cluster.  The tendency of segments of similar spectral
character to form contiguous regions gives us confidence in the
method.  Summed spectra of all segments in several clusters are shown
in Figure~\ref{spectra}.  We shall refer to the clusters by
abbreviations for their colors in Fig.~\ref{ims1}: R (red), O
(orange), Y (yellow), LG (light green), DG (dark green), LB (light
blue), DB (dark blue), P (purple) and B (black).  The values of the
count ratios that define the center of each cluster are given in
Table~\ref{centers}.

The clusters containing the most prominent emission in the
ejecta-dominated bands are regions LB, DG, LG, and Y. Figure 2 (top)
shows region LG contrasted with CSM-dominated regions, while the
others are shown in Figure 2 (bottom).  Each of their spectra
contained similar Fe L, Si K$\alpha$ and K$\beta$, S K$\alpha$ and
K$\beta$, Ar K$\alpha$, and Ca K$\alpha$ features with the exception
of region DG, whose features were much more muted.  Region LG is the
brightest of the four, because of its close proximity to the bright
north rim. Region Y contains the highest concentration of iron in the
remnant in a localized patch. The outer ejecta knots containing
lighter materials are contained primarily in region O as seen by the
very high silicon to iron ratios within the spectrum. Synchrotron
radiation is represented by region P as the spectrum is almost
featureless. Regions R, DB, and B contain the clearest Mg feature,
suggesting CSM.  These regions are compared in the two spectral plots
of Figure 2. The vast majority of the CSM is contained within region
R. Confirmation of this separation's effectiveness is evidenced by
comparing region R to the {\sl Spitzer} 24 $\mu$m image of SN1604
(Fig.~\ref{csmim}; Blair et al.~2007) which highlights the heated dust
that is evidence of CSM. The two images are extremely similar.

We can examine this issue more closely.  Figure~\ref{spectra} shows
the integrated spectrum of region R (red, top), along with region LG
(green, second from top), west-central portion of region R (third from
top) with local background subtracted, and finally the spectrum of a
small knot reproduced from Reynolds et al.~(2007), also with a local
background subtracted (bottom).  The west-central spectrum also shows
the CSM component from a multicomponent fit (dotted line; see below).
The contrast between regions R and LG is evident: the different peak
energies result from a higher contribution from Ne in region R and
from Fe L in region LG.  Region R also shows a feature between 0.6 and
0.7 keV, O Ly $\alpha$, which is more obvious in the lower two
spectra.  The obviously greater prominence of both O and Mg in all of
Region R (especially in the central emission) and the smaller
contribution from Fe confirm the association of those regions with
CSM.

While we might expect CSM to be found around the outer edge of the
SNR, its presence across the central region is striking.  Blair, Long,
\& Vancura (1991) first pointed out the presence of nonradiative
shocks there, indicating partially neutral upstream material.  They
found that knots in the eastern part of the central emission showed
redshifts of order 500 km s$^{-1}$ while those in the western part
were blueshifted by about $-600$ km s$^{-1}$, indicating that the
former were on the far side and the latter on the near side, projected
against the center.  Our completely independent analysis locates the
same regions, and we adopt the same interpretation.

\section{Quantitative analysis of CSM}

\begin{figure}
\centering
%\epsscale{1.15}
\epsscale{1.1}
%\plotone{f1p.png}
\includegraphics[width=3truein]{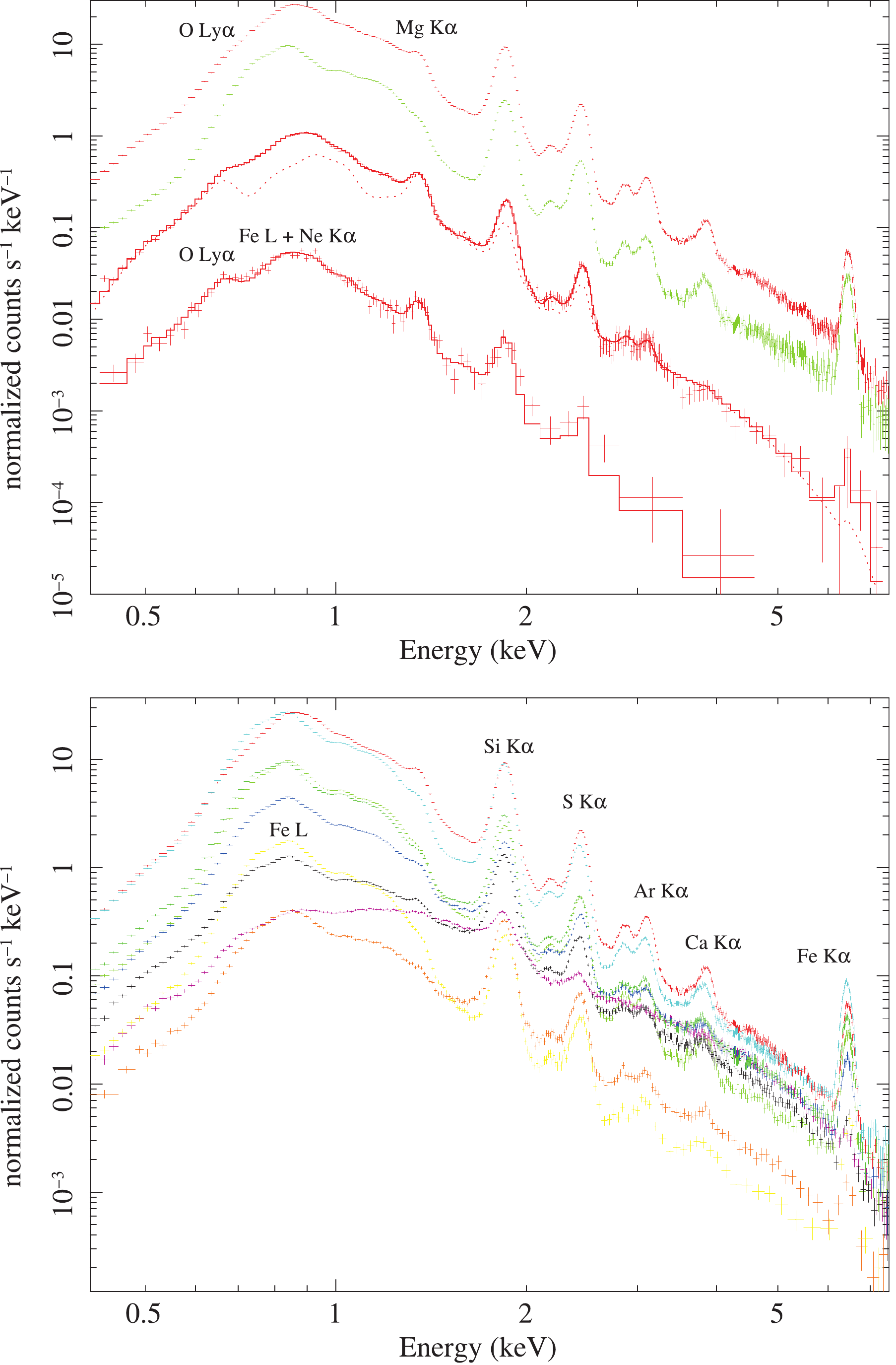}
\caption{Top: Spectra of primary CSM regions compared to
ejecta-dominated Region LG.  Top: Total of Region R.  Next: Region LG.
Third: West central part of Region R, with local background
subtracted; dotted curve is the CSM component from the multicomponent
spectral fit described in the text.  Bottom: Small knot in central
region with local background subtracted, reproduced from Reynolds et
al.~(2007).  Notice the clear O and Mg features in all three CSM
spectra.  Bottom: Spectra of primary ejecta regions compared to the
synchrotron region and Region R. Comparable in strength to Region R is
Region LB; note the difference in shape near the peak, due to Fe L
dominance of Region LB.  In order of decreasing brightness at 1 keV:
Regions LG, DG, DB, Y, B, P (synchrotron) and O.}
%\vspace{-0.1truein}
\label{spectra}
\end{figure}

The integrated spectrum of Region R clearly shows distinctive features
consistent with its identification as CSM.  Ideally, we would use
oxygen as a tracer of CSM, as relatively little O is present in
most SN Ia models (e.g., Maeda et al.~2010).  However, while an
inflection at the energy of O Ly$\alpha$ (0.65 keV) can be seen in
some of the spectra of Figure~\ref{spectra}, the substantial absorbing
column density makes quantitative analysis difficult.  The clearest
signal of an element not expected to be synthesized in much quantity
in a SN Ia (e.g., Maeda et al.~2010) comes from Mg K$\alpha$ (1.34
keV), with the absence of Mg Ly$\alpha$ (1.47 keV) as a constraint.
The regions we identify as CSM have weak, but well-isolated and easily
characterized, emission features for Mg K$\alpha$, and we use that
line to quantify CSM (Figure 3, center).

\begin{figure}
\centering
%\epsscale{1.15}
\epsscale{1.1}
%\plotone{f1p.png}
\includegraphics[width=2.9truein]{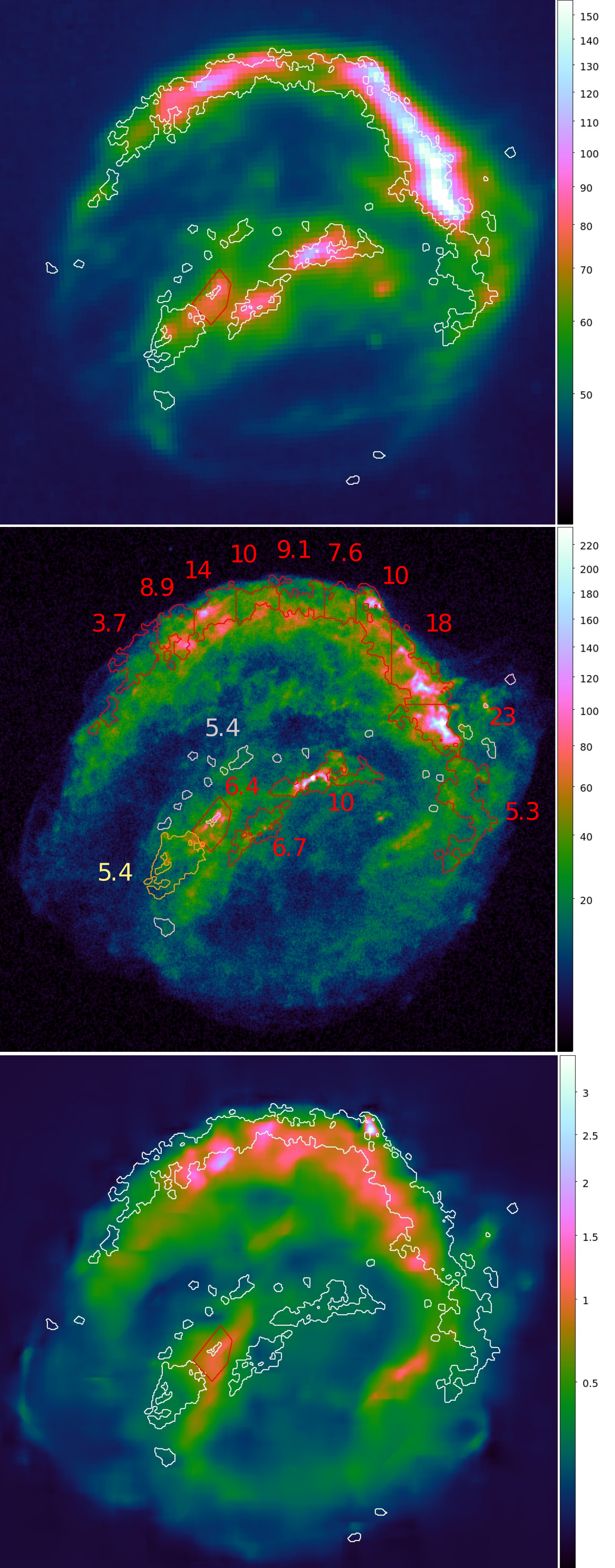}
\caption{Top: Region R (contours) plus an IR-bright central subregion
of Region LG superposed on the 24 $\mu$m {\sl Spitzer} image (Blair et
al.~2007).  Color scale is in MJy sr$^{-1}$. Middle: Soft X-ray image
(ct pixel$^{-1}$ in 0.3 -- 0.72 keV) from {\sl Chandra}, with contours
of subregions of CSM-dominated Region R (contours) plus the IR-bright
central LG subregion.  Numbers: Mg line-strength surface brightnesses
(units: $10^{-8}$ photons cm$^{-2}$ s$^{-1}$ arcsec $^{-2}$).  Bottom:
Region R contours superposed on a smoothed image from 6.2 to 6.8 keV
(Fe K). The IR-bright LG subregion overlaps with the peak of central
Fe K emission. Note the poor correlation between strong Fe emission
and CSM, especially in the west-central region.
}
%\vspace{-0.1truein}
\label{csmim}
\end{figure}

We subdivided the CSM-dominated region R into 14 subregions, and
extracted spectra from each.  We subtracted background from a large
region to the north of the remnant.  Spectra were fitted with two
Gaussians for the two lines and power-law model for the continuum,
between 1.2 and 1.6 keV, yielding line strengths.  The Ly$\alpha$
feature was often negligible.  For each subregion, we calculated the
line-strength surface brightness (ph cm$^{-2}$ s$^{-1}$
arcsec$^{-2}$).  These values are shown in Figure 3 (center).

The west-central portion of region R, with a surface area of 410
arcsec$^2$ and a Mg K$\alpha$ flux of $4.2 \times 10^{-5}$ ph
cm$^{-2}$ s$^{-1}$, rivals the bright northern rim in Mg K$\alpha$
surface brightness. In order to learn more about its plasma
properties, we modeled its spectrum with a simple plane shock ({\tt
vpshock} model in XSPEC). As in Reynolds et al.~(2007), we assumed an
absorbing column density $N_H = 5.2 \times 10^{21}$ cm$^{-2}$. Wilms
et al.~(2000) solar abundances were adopted for all elements except
for N (fixed to $3 \times$ solar) and Ne and Mg (that we allowed to
vary in the fitting process). The ejecta contribution was modeled by a
pure Fe, NEI v1.1 {\tt vpshock} model plus a separate
single-ionization-timescale model {\tt vnei} containing
intermediate-mass (Si, S, Ar, and Ca) elements. (We also added a
Gaussian line at 0.73 keV in order to account for missing Fe lines in
the NEI v1.1 atomic code.) Ejecta contribute to the spectrum mostly in
Fe L- and K-shell lines, and in K$\alpha$ lines of Si, S, and Ar (see
Fig.~2).  The temperature and ionization age of the dominant CSM
component are 1.2 keV and $1.1 \times 10^{11}$ cm$^{-3}$ s, and the
fitted Ne and Mg abundances are near solar (0.7 and 1.2,
respectively).  This component is plotted as the dotted curve in
Figure 2; it accounts for almost all the emission except for obvious
Fe L and K emission.  Fits with a more elaborate {\tt vnpshock} model
with unequal ion and electron temperatures reproduce these results.
For an assumed shock velocity of $\sim 1500$ km s$^{-1}$
(corresponding to the mean temperature of 2.7 keV), typical for
Balmer-dominated shocks in Kepler's SNR (Blair et al.~1991), electrons
are heated first to 0.8 keV at the shock front, and then gain energy
through Coulomb collision with ions. Average properties of the shocked
CSM in the west-central portion of region R appear typical of Kepler
as a whole, although its emission measure of $0.32 M_\odot d_4^2$
cm$^{-3}$ is only a small fraction of the total emission measure of
$\sim 10 M_\odot d_4^2$ cm$^{-3}$ (see Blair et al.~2007 for a
discussion of the CSM plasma properties as derived from the
spatially-integrated {\it XMM-Newton} RGS spectrum). Blue-shifted
optical emission found at this location by Blair et al.~(1991)
indicates that this material was expelled toward us by the SN
progenitor.  We also performed a spectral analysis of the adjacent
central portion of region R farther south and east that contains
red-shifted optical emission from material expelled away from us by
the SN progenitor. We found the same Ne and Mg abundances there but a
slightly higher temperature of 1.5 keV, a somewhat shorter ionization
age of $7.1 \times 10^{10}$ cm$^{-3}$ s, and a significantly smaller
emission measure of $0.078 M_\odot d_4^2$ cm$^{-3}$. These results may
be interpreted as a modest density effect, where the blue-shifted
west-central portion of the CSM is somewhat denser than the
red-shifted portion farther south and east.

Although the spatial correlation between region R and the IR emission
is reasonably good, it is far from perfect. One reason for this might
be superposition of ejecta and CSM along the line of sight. We
examined X-ray spectra at an IR bright east-central location (see
Fig.~\ref{csmim}) that is part of the ejecta-dominated LG
region. While the Fe L-shell emission dominates, a strong Mg K$\alpha$
line is also present, with a surface brightness as high as found in
central portions of region R. An image in the 6.2 -- 6.8 keV energy
range shows a prominent Fe K$\alpha$-emitting filament intersecting
the central CSM band at this location (Fig.~\ref{csmim}).  Apparently,
both the CSM and Fe-rich ejecta contribute significantly to the X-ray
spectrum of this region. They may even arise from physically adjacent
regions, as the collision of Fe-rich ejecta with denser than average CSM
might be expected to enhance Fe L- and K-shell emission. But the
overall spatial correlation between Fe-rich ejecta (as traced by the
Fe K$\alpha$ emission) and the CSM is quite poor in the central
regions of the remnant, presumably because of the asymmetric
distribution of Fe within the SN ejecta.

\begin{figure}
\centering
%\epsscale{1.15}
\epsscale{1.1}
%\plotone{f1p.png}
\includegraphics[width=2.8truein]{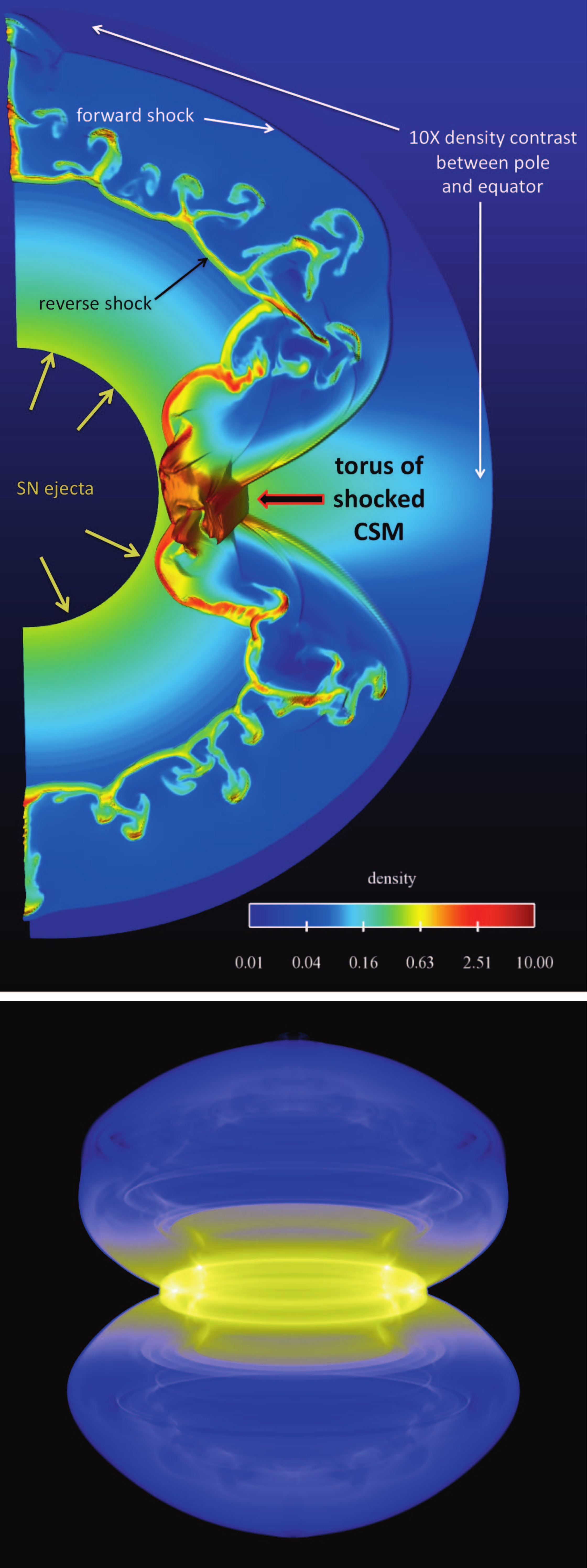}
\caption{Top: Hydrodynamic simulation of density as a blast wave
encounters an equatorial wind.  At this time, the swept-up mass is
comparable to the ejected mass of $1.4 M_\odot$.  The wind speed has
been neglected compared to the blast-wave speed.  Bottom:
Line-of-sight integration of density squared from the hydrodynamic
simulation.  The symmetry axis is tilted out of the sky plane by
10$^\circ$.  }
%\vspace{-0.1truein}
\label{hydro}
\end{figure}

\section{Hydrodynamic simulations}

The presence of the band of CSM across the center of Kepler (echoed by
the presence of nonradiative H$\alpha$ emission) indicates that this
material is seen in projection in front and in back of the remnant.
Such a morphology could result from a pre-SN CSM distribution that is
predominantly disk-like, as expected for mass loss from AGB stars (see
Section 6 below for references), with the line of sight roughly in the
plane of the disk.  To examine this possibility more closely, we
performed 2-D hydrodynamic simulations of an ejecta-driven blast wave
expanding into an azimuthally varying stellar wind, with density
varying by a factor of 10 from pole to equator.  Half the wind mass is
within $10^\circ$ of the equator.  We ignored the wind speed as it is
a few tens of km s$^{-1}$, negligible compared to the blast-wave
speed.  We did not attempt to model the north-south density gradient
(Blair et al.~2007), presumably the result of system motion to the
north or northwest (Bandiera 1987). We modeled the SN ejecta with an
exponential profile appropriate for Type Ia explosions (Dwarkadas \&
Chevalier 1998).  We used the well-tested code VH-1, a conservative,
finite-volume code for evolving the Euler equations describing an
ideal, compressible gas.  The details of the numerical simulation
follow the procedure described in Warren \& Blondin (2012).

Figure~\ref{hydro} shows a stage at which the swept-up CSM mass is
about equal to the ejected mass ($1.4 M_\odot$).  A torus of shocked
CSM occupies the equatorial plane.  The lower panel of
Figure~\ref{hydro} shows a 3-D projection of the model, integrating
the square of the density (proportional to emission measure) along
lines of sight, with the symmetry axis tilted by $10^\circ$ to the
plane of the sky.  The result is an incomplete bar of shocked CSM
across the remnant center as observed, whose radius is about half that
of the extent to the north and south, in rough agreement with the
central bar of CSM we observe.  While the simulation is only
suggestive, it indicates that more detailed study of such a model may
lead to an improved understanding of Kepler's dynamics.

\section{Discussion}

Our observations of CSM toward the center of Kepler are naturally
explained by a disk seen edge-on, while the bright northern rim results from
northward motion of the system (Bandiera 1987).  Such an asymmetric
distribution of CSM is expected from a binary system, where an
equatorial disk of enhanced mass loss is likely.  Inferences of CSM
around at least a few Type Ia SNe continue to accumulate (e.g., SN
2008J; Taddia et al.~2012; PTF11kx; Dilday et al.~2012).  In
particular, Dilday et al.~(2012) infer a substantially asymmetric CSM
distribution, suggesting a symbiotic binary progenitor system with
mass loss concentrated in the orbital plane.  In an SD model for
Kepler, as required by the CSM, we expect the companion to have been
an AGB star.  Some AGB stars show dense envelopes accessible to study
through molecular emission such as CO; on scales of $10'' - 20''$, the
emission is relatively symmetric (Neri et al.~1998).  A few systems
seem to require asymmetry, though the incomplete sampling of the
3-element IRAM interferometer used by Neri et al.~(1998) make detailed
imaging impossible.  Additional observational evidence for asymmetric
winds from evolved stars is presented by Chiu et al.~(2009) and
Huggins (2007).  On theoretical grounds, we expect that the winds from
AGB stars in detached binaries can, through gravitational focusing by
the companion, produce highly asymmetric CSM, with density contrasts
of 10 or more found in numerical simulations by Mastrodemos \& Morris
(1999) and characterized by Huggins, Mauron, \& Wirth (2009).
Politano \& Taam (2011) find that several percent of AGB systems
should show such strong asymmetries.  In a particular case, 3D
simulations of RS Oph (Walder, Folini, \& Shore 2008) show very large
equator-to-pole density variations on the scale of the orbital
separation, averaging to factors of 2 -- 3 on much larger scales.
These simulations do not consider the possibility of a wind from the
white-dwarf companion (Hachisu et al.~1996), which would tend to
evacuate material perpendicular to the plane of the disk and enhance
the equator-to-pole density contrast.

The distribution of strong Fe emission in Kepler is worthy of note.
Most Type Ia SN models produce highly stratified ejecta, so most Fe
should be in the remnant interior.  Where the ejecta impact the dense
wind in the equatorial plane, then, one might expect enhanced Fe
emission -- so in a plane roughly coincident with the plane of central
CSM.  This does not appear to be the case.  Fe K$\alpha$ emission
toward Kepler's interior is asymmetrically distributed, with one patch
near the east-central CSM emission, but less near the west-central
CSM.  We speculate that one cause of Fe asymmetry might be the
``shadow'' in Fe cast by the companion star, blocking the ejection of
material in that direction.  Pan et al.~(2012) estimate that a RG
companion can shadow up to 18\% of the solid angle from the ejecta of
a Type Ia SN, while Garc\'{i}a-Senz et al.~(2012) show that
morphological effects of a companion's shadow can survive for hundreds
of years.  Further study of Fe in particular in Kepler will allow us
to examine this possibility in more detail.

\section{Conclusions}

We have used Gaussian mixture decompositions of multicolor X-ray
spectral data from the long {\sl Chandra} observation of Kepler's
supernova remnant to identify and characterize regions of shocked
circumstellar material, as distinct from the ejecta that dominate the
integrated spectrum.  We find that shocked CSM is co-located with both
nonradiative shocks identified by H$\alpha$ emission, and with 24
$\mu$m dust continuum emission seen with {\sl Spitzer}.  We suggest
that the central band of CSM is the remnant of a circumstellar disk
seen edge-on.  Our 2-D hydrodynamic simulation shows that a blast wave
encountering an equatorial wind from a companion can naturally produce
this morphology.  The asymmetry we observe requires a binary
progenitor system in which the donor is an evolved (AGB) star.

\acknowledgments
This work was supported by the National Science Foundation through
award AST-0708224 and by NASA through grant NNX11AB14G.

\clearpage

\begin{deluxetable}{l r r r r r r r r r}
\tablecolumns{10} \tablewidth{0pc} \tabletypesize{\footnotesize}
\tablecaption{Cluster Centers}

\tablehead{ \colhead {Energy Band} & B & P & DB & LB & DG & LG & Y & O & R}

\startdata
O    & 0.89 & 0.73 & 1.01 & 1.13 & 1.1  & 1.11 & 1.23 & 0.87 & 1.05 \\
Fe   & 1.2  & 1.07 & 1.35 & 1.49 & 1.44 & 1.47 & 1.68 & 1.31 & 1.39 \\
Mg   & 0.9  & 0.87 & 0.96 & 1.02 & 0.97 & 1.03 & 1.07 & 0.89 & 1.05 \\
Si/S & 1.1  & 0.97 & 1.11 & 1.19 & 1.16 & 1.17 & 1.19 & 1.13 & 1.14 \\
\enddata

\tablecomments{Entries are the logarithms of ratios of counts in the
four energy bands to continuum at the centers of each of the nine
clusters identified by the Gaussian Mixture Method.  Bands are
described more fully in the text.}
\label{centers}
\end{deluxetable}

\end{document}